\documentclass[12pt]{article}
\usepackage[dvips]{graphics}
\usepackage[pdftex]{graphicx}
\usepackage{graphicx}
\usepackage{color}
\usepackage{epsfig}
\usepackage{latexsym}
\usepackage{amssymb}
\usepackage{color}
\usepackage{amsmath}
\usepackage{float}
\usepackage{mathrsfs}
\usepackage{color}
\usepackage{subcaption}

\newcommand{\kom}[1]{}
\renewcommand{\kom}[1]{{\bf [#1]}}
\definecolor{blau}{rgb}{0.1,0.0,0.9}

\newcounter{komcounter}
\numberwithin{komcounter}{section}

\newtheorem{remark}{Remark}
\newtheorem{algorithm}{Algorithm}
\newtheorem{example}{Example}

\title{Enhancing Precision with the Local Pivotal Method:\\ A General Variance Reduction Approach 
}

\author{
Marcus Olofsson{\small{$^1$}},
Anton Grafstr\"om{\small{$^2$}},
Niklas L. P. Lundstr\"{o}m{\small{$^1$}}
\\\\
{\small{$^1$}}\it \small Department of Mathematics and Mathematical Statistics, \\
\it \small Ume{\aa} University, SE-90187 Ume{\aa}, Sweden\\
\it \small niklas.lundstrom@umu.se, \it \small marcus.olofsson@umu.se
\\\\
{\small{$^2$}}\it \small  Department of Forest Resource Management, \\
\it \small Swedish University of Agricultural Sciences, Ume{\aa}, Sweden \\
\it \small anton.grafstrom@slu.se
}

\date{\today}

\begin{document}


\maketitle


\begin{abstract}
\noindent The local pivotal method (LPM) is a successful sampling method for taking well-spread samples from discrete populations.
We show how the LPM can be utilized to sample from arbitrary continuous distributions and thereby give powerful variance reduction in general cases. The method creates an ``automatic stratification" on any continuous distribution, of any dimension, and selects a ``thin" well-spread sample. We demonstrate the simplicity, generality  and effectiveness of the LPM with various examples, including Monte Carlo estimation of integrals, option pricing and stability estimation in non-linear dynamical systems. Additionally, we show how the LPM can be combined with other variance reduction techniques, such as importance sampling, to achieve even greater variance reduction. To facilitate the implementation of the LPM, we provide a quick start guide to using LPM in MATLAB and R, which includes sample code demonstrating how to achieve variance reduction with just a few lines of code.
\end{abstract}


\section{Introduction} \label{sec:introduction}

Let's assume we wish to estimate the mean $\mu$ of some trait in a discrete population $U=\{1,2,...,N\}$. A classical estimator is the Horvitz-Thompson estimator, see \cite{HT52}
\begin{equation} \label{eq:HTestimator}
\hat \mu_{HT}:= \hat \mu := \frac{1}{N} \sum_{i\in S} \frac{y_i}{\pi_i},
\end{equation}
where $S\subset U$ is a sample of size $n$, $y_i$ is the value of the trait for individual $i$, and $\pi_i$ is the inclusion probability of individual $i$ given some sampling design. For equal inclusion probabilities $\pi_i=n/N$ this simply corresponds to the sample average. The random sample $S$ should preferably be such that $\hat \mu$ lies close to the true value $\mu$ with small variance in the set of allowed samples.  If the sample size can be arbitrarily large and the sample is without replacement, this is trivially achieved as taking $n=N$ gives the true value $\mu$ with $0$ variance. However, if the cost of collecting or retrieving the values $y_i$ is high, one needs to take care when choosing the sample so that the estimator \eqref{eq:HTestimator} gives an acceptable result with small variance also for modest values of $n$. This is the motivation for \textit{variance reduction} techniques.

Variance reduction is an indispensable necessity in many branches working with statistical estimation. Classical variance reduction methods include control variates, antithetic variables, stratified sampling and importance sampling, see e.g. \cite{G04}. Stratified sampling is perhaps one of the most used due to its simplicity and efficiency. The main idea is to ensure a well-spread and proportional sample by dividing the space of outcomes into smaller subsections, \textit{"stratas"}, and generate each sample point conditional on being in a given stratum. A sample thereby spreads over the entire population by design. In general, it is well established that such well-spread samples are efficient and favourable to sampling with independent observations. 
A drawback of stratified sampling is that it may be non-trivial to construct appropriate strata and to allow for unequal inclusion probabilities, especially if the number of auxiliary variables is large.

Building on \cite{DT98} the authors provide in \cite{GLS12} a novel sampling method, the \textit{local pivotal method} (LPM), to select a well-spread sample consisting of a small fraction of a discrete population. Roughly, the LPM creates an automatic stratification and selects a ``thin" well-spread sample from the population, while respecting possibly unequal inclusion probabilities. In several applications where multiple auxiliary variables are available and the cost of collecting observations is relatively high, the LPM has successfully been used to reduce variance. Environmental inventories and inventories of forests and landscapes, which require expensive field visits to measure target variables, apply the LPM. For example, the Swedish national forest inventory and the National Inventory of Landscapes in Sweden use this form of sampling, see \cite{Gea17a} and \cite{AKB23}. The LPM has also found applications in agricultural surveys, e.g. \cite{BPP15}, and soil surveys, e.g. \cite{B21}, 
and has been used for sampling from continuous populations \cite{GM18, Gea17b}. 

Although the LPM is flexible, easy to implement (also with unequal inclusion probabilities), and has proven to be remarkably efficient, it has not yet been adopted by a broader community. This is the main motivation for the current paper, in which we present the fundamentals of LPM and demonstrate its applicability to arbitrary continuous distributions of any dimension. %
%
%
With examples from Monte Carlo estimation of integrals, pricing of European options and estimation of stability for a rain forest and for the rotor in a hydro power generator, we show how to use LPM to achieve fast and reliable variance reduction with minimal effort in a wide range of applications.

An R-library named \texttt{BalancedSampling}, see \cite{GLP22}, and a MATLAB implementation are readily available. The paper is concluded with a quick start guide on how to use these implementations.



\section{LPM for continuous distributions}

Motivated by applications in forestry, the LPM was originally designed for taking well-spread samples of discrete populations. We begin by describing this procedure.

Let $U$ denote a discrete population consisting of $N$ units, each with a prescribed inclusion probability $\pi_i$, possibly given by a function of some auxiliary measurable variable $z_i$. The LPM selects a sample from $U$ stepwise, by updating the inclusion probabilities of the population so that the sampling outcome is decided for at least one of two neighbouring units in each step. More explicitly, given two neighbouring units $i$ and $j$, we randomly update their probabilities $(\pi_i, \pi_j)$ to
\begin{align} \label{eq:updatedprob}
&(\pi'_i,\pi'_j)= \begin{cases} (0,\pi_i+\pi_j) & \mbox{with probability $\frac{\pi_j}{\pi_i+\pi_j}$} \\
(\pi_i+\pi_j,0) & \mbox{with probability $\frac{\pi_i}{\pi_i+\pi_j}$}
\end{cases} & \mbox{\qquad if $\pi_i+\pi_j <1$}, \notag  \\
&(\pi'_i,\pi'_j)= \begin{cases} (1,\pi_i+\pi_j-1) & \mbox{with probability $\frac{1-\pi_j}{2-\pi_i-\pi_j}$} \\
 (\pi_i+\pi_j-1,1) & \mbox{with probability $\frac{1-\pi_i}{2-\pi_i-\pi_j}$}
\end{cases} & \mbox{\qquad if $\pi_i+\pi_j \geq 1$.}
\end{align}
In other words, we move inclusion probability mass from one unit to the other,
so that either the receiving unit is included in the sample or the giving unit is excluded.
This procedure is repeated until all units have (updated) inclusion probability $0$ or $1$ and the corresponding LPM-sample $S \subset U$ is then given by the units with (updated) inclusion probability $1$. The sample $S$ is well spread over the population and has expected sample size $n= \sum_i \pi_i$. Provided that $n$ is integer, the sample size is fixed.\footnote{It is not necessary that $n$ is an integer, but we will stick to this case for simplicity in the presentation.}


In detail, the LPM operates as follows on a discrete population:
\begin{algorithm}[The local pivotal method for selecting a sample] \label{algorithm}
\begin{enumerate}
\item[]
\item[$i)$] \mbox{Randomly select unit $i$ among those with updated probability strictly between 0 and 1.}
\item[$ii)$] \mbox{Find a nearest neighbour $j$ to $i$.}
\item[$iii)$] \mbox{Update the probabilities for units $i$ and $j$ according to \eqref{eq:updatedprob}}.
\item[$iv)$] \mbox{Repeat from $i)$ until all units have an updated probability equal to $0$ or $1$.}
\end{enumerate}
\end{algorithm}

\begin{remark}
At the cost of some additional computations one can achieve an even more well spread sample by replacing $iii)$ in the algorithm above with
\begin{enumerate}
\item[$iii')$] \mbox{If $j$ has $i$ as its nearest neighbour, then update the probabilities according to \eqref{eq:updatedprob}.}
\end{enumerate}
The procedure in Algorithm \ref{algorithm} is typically sufficient for practical applications.
When relevant, we denote by "LPM2"  ("LPM1") the algorithm using $iii$ ($iii'$), but simply write LPM when discussing the method in general.
\end{remark}


\begin{remark}
Note that any distance function $d(i,j)$ is allowed when determining the "nearest neighbour" in step $ii)$. In particular, the distance can be measured in an arbitrary space, allowing the user to find well spread samples in an auxiliary space consisting of a large number of variables. 
\end{remark}

\subsection{Extension to continuous distributions}

In order to apply LPM on a continuous population having distribution $\tilde U$, we only need to add one single step to Algorithm \ref{algorithm} -- a discretization of the population.
\begin{enumerate}
\item[$\emptyset)$] \mbox{\textit{Draw $N$ independent points from $\tilde U$.}}
\end{enumerate}

\noindent
We can now apply steps \emph{i)--iv)} in Algorithm \ref{algorithm} to the discrete subset produced in step $\emptyset$ to get a LPM-sample of the continuous population $\tilde U$. The discretization in step $\emptyset)$ does not introduce any bias.

When applying LPM to continuous distributions, one must choose both the sample size $n$ and the discretization size $N$. There is no definitive answer as to how large $N$ should or needs to be, but our examples below indicate that $N = 10 \cdot n$ is sufficient for substantial variance reduction and that no more than $N=100 \cdot n$ is needed to get most of the benefits of LPM, see Figure \ref{fig:decrease}. If $n=N$, then LPM coincides with independent identically distributed (iid) observations, and no reduction in variance can be achieved.  We remind the reader that trait-evaluation is only needed for the $n$ points chosen by the LPM-algorithm and thus increasing $N$ can typically be done at little cost.

We proceed by demonstrating the simplicity and the efficiency of LPM on continuous distributions through examples.

\begin{figure}
\begin{subfigure}{.49\textwidth}
\centering
\includegraphics[height=0.75\linewidth, width=1\linewidth]{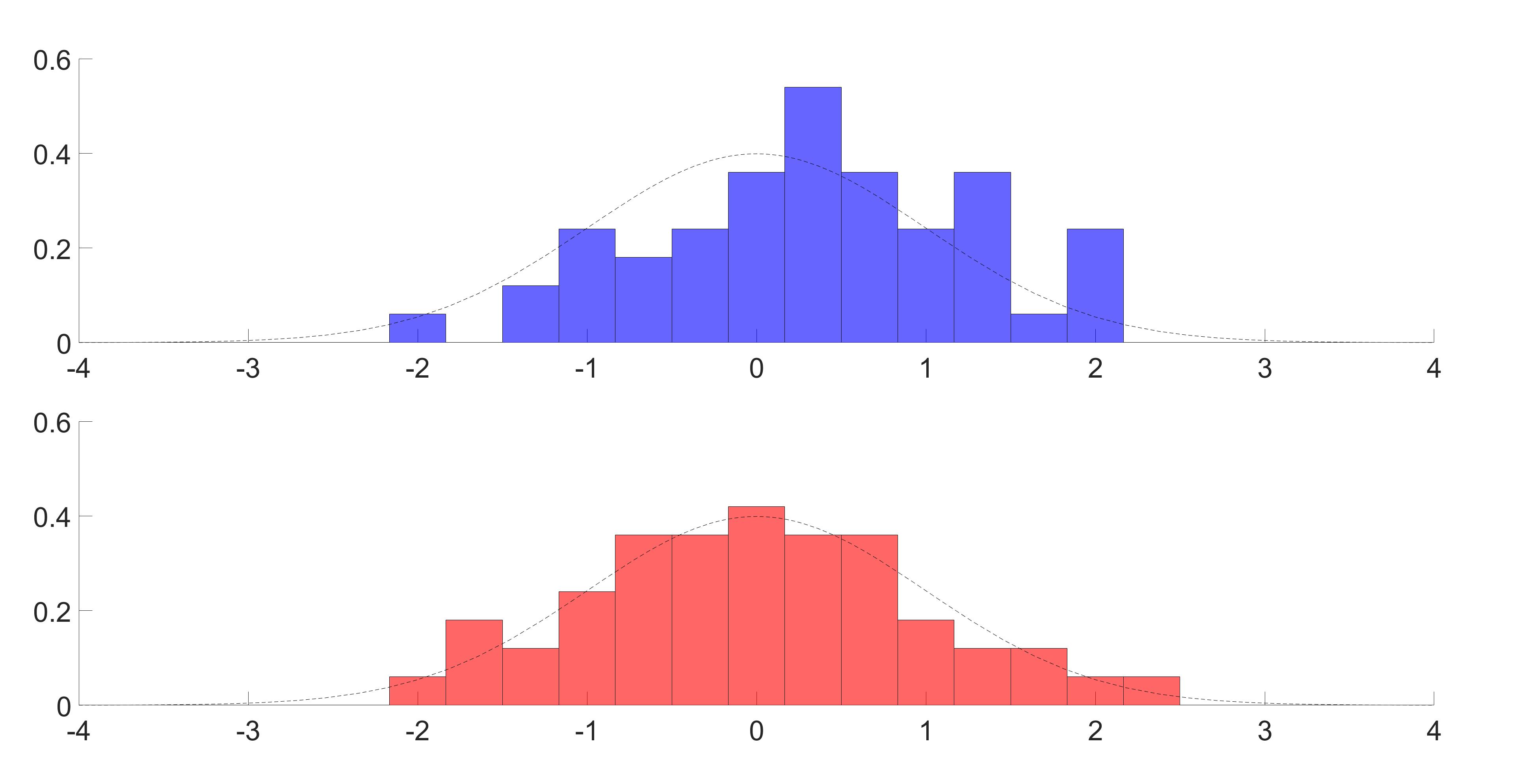}
\caption{$n=50$}
\label{fig:normal50}
\end{subfigure} \hfill
\begin{subfigure}{.49\textwidth}
 \centering
\includegraphics[height=0.75\linewidth, width=1\linewidth]{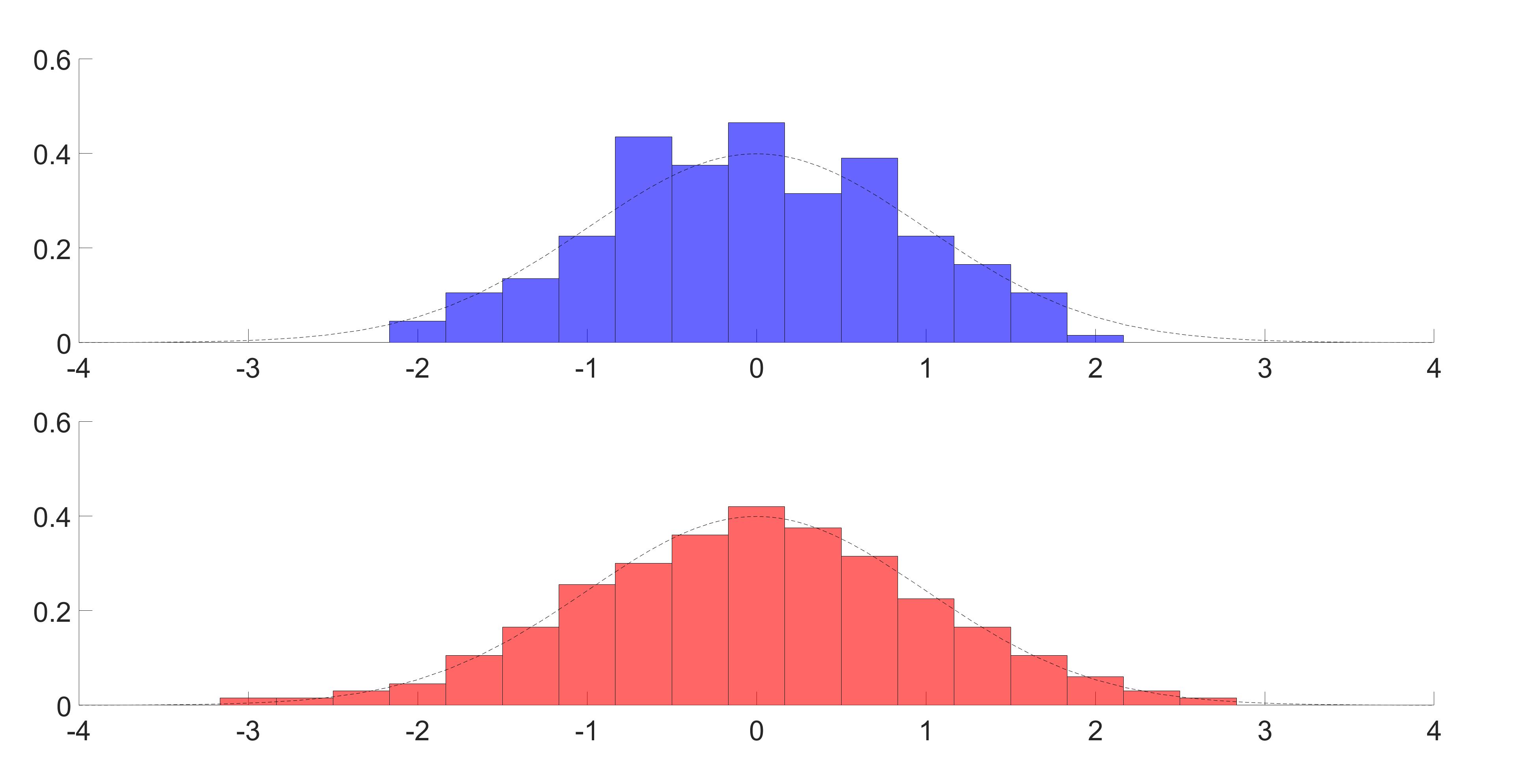}
\caption{$n=200$}
\label{fig:normal200}
\end{subfigure}
\caption{A visual comparison of iid observations (blue, top row) and an LPM (red, bottom row) sample from the standard normal distribution.}
\label{fig:histogrammus}
\end{figure}

\begin{example}[Well-spread samples from the normal distribution.]
We sample $N=10^4$ iid points from the normal distribution and then apply {LPM2} to this subset. The result is a well-spread sample from the normal distribution. A histogram of the produced LPM samples for $n=50$ and $n=200$ points are shown in Figure~\ref{fig:normal50} and \ref{fig:normal200}. For visual comparison we also provide histograms of iid observations with the same sample sizes.
\end{example}

\begin{figure}
\begin{subfigure}{.49\textwidth}
  \centering
  \includegraphics[width=0.7\linewidth]{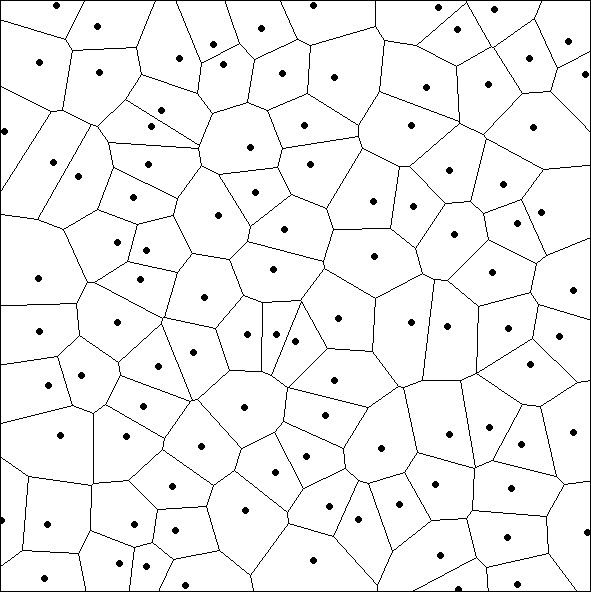}
  \caption{LPM sample.}
\end{subfigure}
\hfill
\begin{subfigure}{.49\textwidth}
  \centering
  \includegraphics[width=0.7\linewidth]{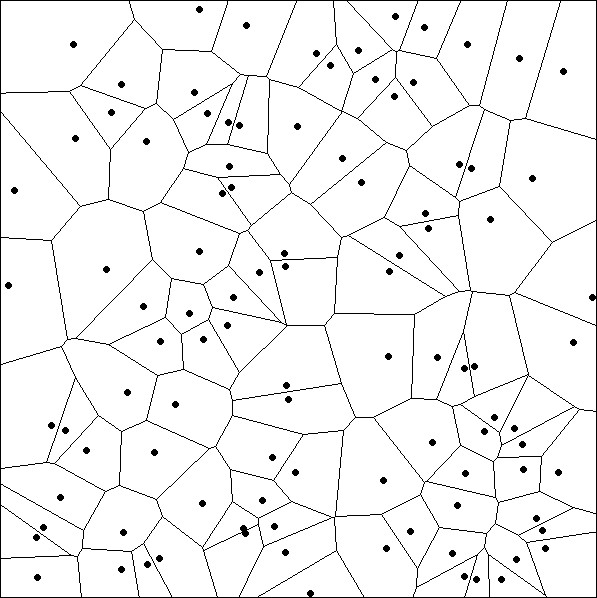}
  \caption{Iid numbers.}
\end{subfigure}
\caption{Voronoi polygons of $100$ LPM and iid 2d uniformly distributed numbers, respectively.}
\label{fig:2Dunif}
\end{figure}

\begin{example}[Well-spread samples from the 2d uniform distribution.]
Figure~\ref{fig:2Dunif} shows Voronoi polygons for a LPM2-sample of $n=100$ observations and $100$ iid observations from the standard $2d$-uniform distribution, respectively. The spatial balance, defined as $\mathbb E\left [ (1/n)\sum_{i=1}^n(a_i-1)^2 \right]$, where $a_i$ is the area of polygon $i$ in the Voronoi tesselation and $n$ equals the number of points and the total area, is approximately $0.065$ for the LPM tesselation and $0.313$ for the iid tesselation, based on $10^4$ simulations using $n=10^2$ and $N=10^4$.
\end{example}

\begin{figure}
\begin{subfigure}{.49\textwidth}
  \centering
  \includegraphics[width=1\linewidth]{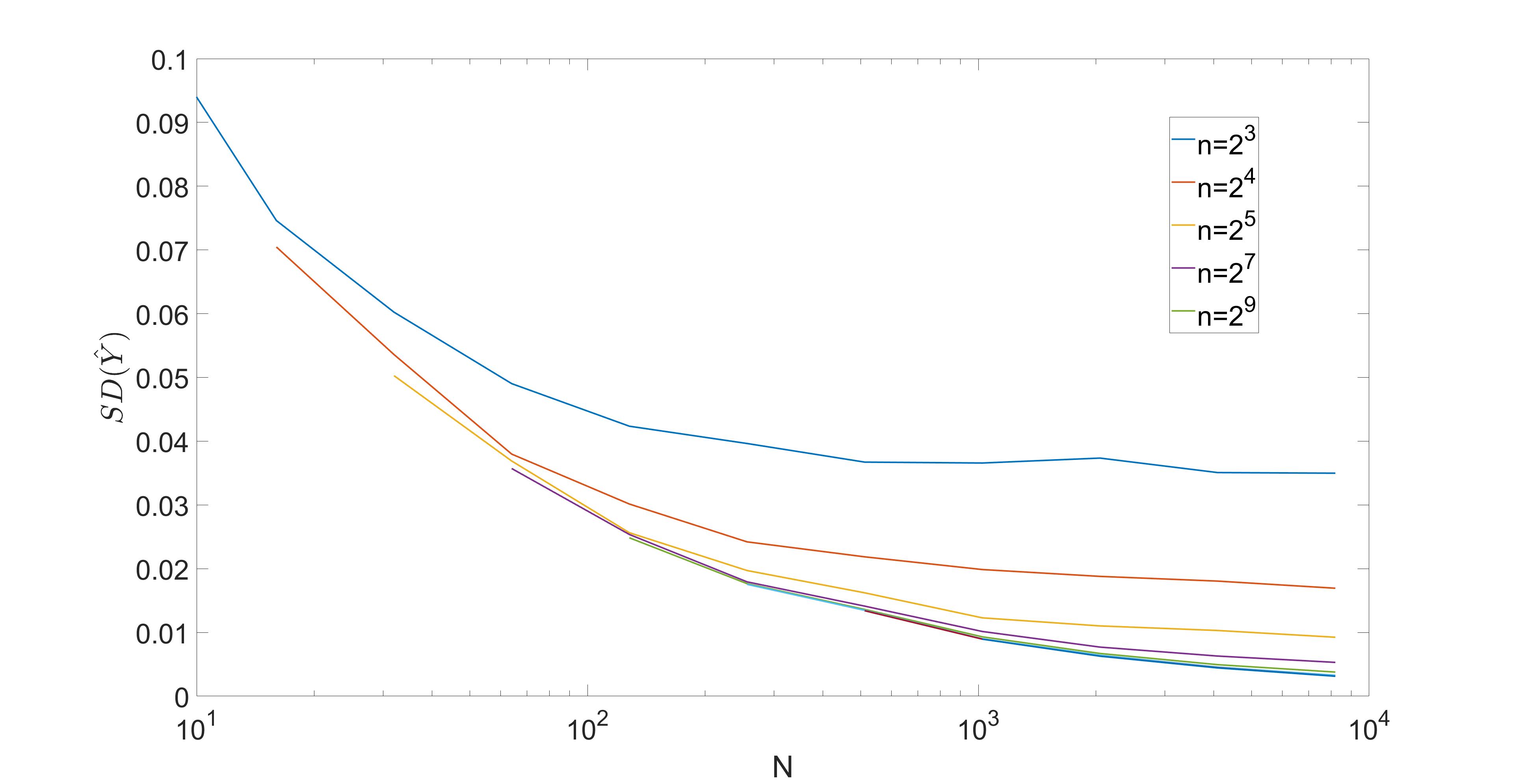}
  \caption{Example \ref{sec:toyexample}, $\mu:=\int_0^1 x dx$. } 
\label{fig:toydecrease}\end{subfigure} \hfill
\begin{subfigure}{.49\textwidth}
  \centering
  \includegraphics[width=1\linewidth]{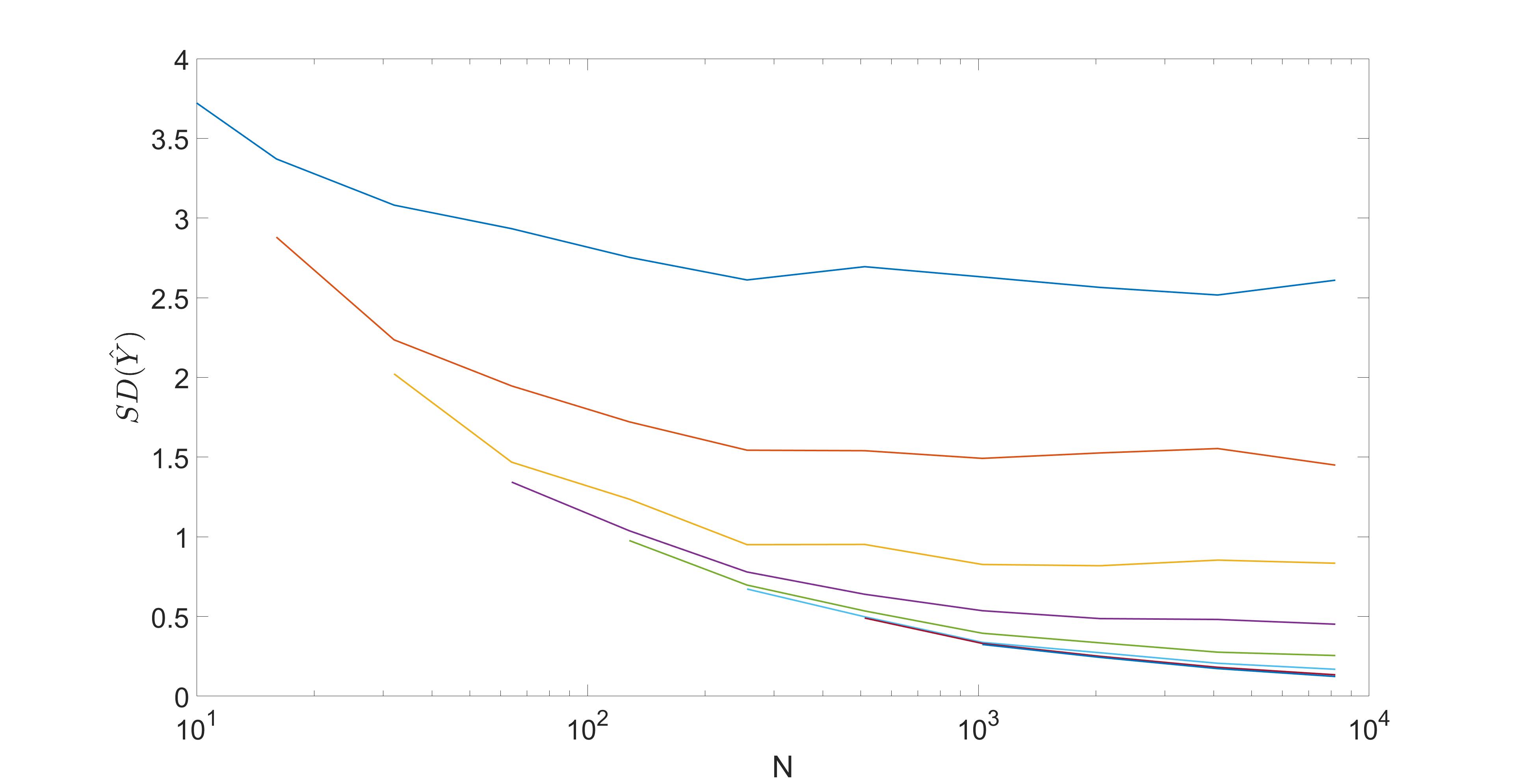}
  \caption{Example Application \ref{sec:EC}, European call option. } 
\label{fig:ECdecrease}\end{subfigure}
\caption{Standard deviation of estimate as a function of the discretization parameter $N$ for varying sample sizes $n\leq N$.} \label{fig:decrease}
\end{figure}

\begin{example}[Monte-Carlo integral estimation.]\label{sec:toyexample}
To estimate the integral
\begin{equation}\label{eq:integral}
\mu:=\int_0^1 x dx = \frac{1}{2},
\end{equation}
the standard Monte-Carlo approach is to take a sample of $n$ iid observations from the uniform $(0,1)$-distribution and calculate the sample average. This works as the integral can be written as the expected value of a function $y(X)=X$, where $X$ has a $U(0,1)$ distribution, i.e. $\mu= \mathbb E\left[y(X)\right]=\int_{0}^{1}y(x)f(x)dx$, where $f$ denotes the uniform probability density function and $f(x)=1$ for $x\in (0,1)$ and 0 otherwise. To draw $n$ uniform observations on $(0,1)$ produces the sampling intensity function $\pi(x)=nf(x)=n$ on $(0,1)$. The estimator can be written as $$\hat \mu =\sum_{x \in S} \frac{y(x)}{\pi(x)}=\frac{1}{n}\sum_{x \in S}y(x),$$ where $S=\{X_1,X_2,\dots,X_n\}$ denotes the random sample of size $n$. By using a well-spread sample rather than independent observations, better estimates can be achieved with smaller samples.

Table~\ref{tab:LPMtoy} shows numerical results when using a LPM2-sample consisting of $10 \%$ or $1 \%$ of a discrete population with size $N=10^4$. LPM halves the standard deviation when using $10$\% of the sample points.
These results are comparable to those achieved by an equally sized sample from $10$ equal and evenly spread strata. However, note that for LPM no effort goes into constructing the strata, something that may require considerable effort in more involved situations. Figure~\ref{fig:toydecrease} shows how the variance of the estimate declines as a function of $N$ for some fixed values sample sizes $n$.

The standard deviations in Table \ref{tab:LPMtoy} for the LPM-estimates and the stratification procedure are found naively by repeating each estimate $m=10^4$ times. We refer to Section~\ref{sec:varianceest} for a discussion on how the variance of the LPM-estimate can be estimated from a single sample.

\begin{table}[h!]
    \centering
    \begin{tabular}{|c|c|c|c|c|c|c|}
\hline
        & $\hat \mu_{iid}$  & $\hat\sigma_{iid}$ &  $\hat \mu_{LPM}$& $\hat \sigma_{LPM}$  & $\hat \mu_{strat}$ & $\hat \sigma_{strat}$  \\ \hline
      $n=10^2$  &  0.525   & 0.028  &  0.508   &  0.004  & 0.496 &   0.003      \\
      $n= 10^3$  & 0.500  & 0.009  &   0.502   &  0.002  &  0.499&   0.001       \\ \hline
    \end{tabular}
    \caption{Numerical estimate of \eqref{eq:integral} using iid uniform random numbers, LPM, and stratification methods. The discretization for LPM is done with $N=10^4$ points.}
    \label{tab:LPMtoy}
\end{table}
\end{example}

\subsection{Combining LPM with other variance reduction techniques}
A major advantage of LPM is its simplicity and flexibility; it can be applied directly to any discrete population~$U$ to create a thinned version of ditto, keeping the main statistical features intact. This allows us to easily combine LPM with other methods for variance reduction, in particular in the context of continuous distributions where we can replace the iid-discretization in step $\emptyset)$ by one based on a variance reduction technique. More explicitly, we can use any suitable variance reduction technique when discretizing the continuous distribution~$\tilde U$ \textit{before} applying LPM to get the variance reduction benefits of both techniques. We exemplify this below by combining LPM with importance sampling.

\begin{example}
Assume that we wish to estimate the expected value $\mu$ of
\begin{equation} \label{eq:Y99}
y_{0.999} (X)=\begin{cases}  0 & \mbox 	{if $X\leq \alpha_{0.999}$} \\
							1000 X &\mbox 	{if $X > \alpha_{0.999}$},
\end{cases}
\end{equation}
where $X \sim \mathcal N(0,1)$ and $\alpha_{0.999}$ is the $99.9$-th percentile of $\mathcal N(0,1)$. This  could be done with an iid sample of size $N$ as
$
\frac{1}{N} \sum_{i=1}^N1000 X_i  \mathbb I \{X_i>\alpha_{0.999}\}
$
where $X_i$ is the value of the $i$-th draw. However, as ${X_i>\alpha_{0.999}}$ is a rare event, a very large sample is necessary to get a good estimate.

A common variance reduction technique when dealing with rare events is \textit{importance sampling}~(IS). The idea of IS is to shift the underlying probability distribution to make the interesting events more likely and then compensate this shift by weighing down the contribution from each sample point. In particular, when estimating $\mathbb E_{\mathbb P} [h(X)]$ under the probability measure $\mathbb P$ with density $f$, we could instead consider $\mathbb  E_{\mathbb Q} \left[h(X) f(X)/g(X)\right]$ under the probability measure $\mathbb Q$ with density $g$, where $f(x)/g(x)$ is the Radon-Nikodym-derivative between $\mathbb P$ and $\mathbb Q$. This is possible since
$$
\mathbb E [h(X)] = \int h(x) f(x) dx= \int h(x) \frac{f(x)}{g(x)} g(x) dx = \mathbb  E_{\mathbb Q} \left[h(X)\frac{ f(X)}{g(X)}\right].
$$
Estimating the latter may give a smaller variance, depending on the choice of $g$. We refer to \cite{G04} for more on IS and how to choose the measure $\mathbb Q$.

More explicitly, if we denote the density for $\mathcal N(0,1)$ by $f$ and that of $\mathcal N(3,1)$ by $g$ and utilize IS to estimate 
$\mu$ we get
$$
\mathbb \mu=E_f [ y_{0.99}(X)] = \int y_{0.99}(x) f(x) dx= \int y_{0.99}(x) \frac{ f(x)}{g(x)} g(x) dx = \mathbb E_g \left[y_{0.99}(X)\frac{ f(X)}{g(X)}\right]
$$
where $\mathbb E_f$ and $\mathbb E_g$ are the expected values under $f$ and $g$, respectively. Taking a sample of iid observations from $\mathcal N(3,1)$ and estimating the last expression with the help of these we find a massive variance reduction, up to $400$ times lower. This iid can then be thinned by LPM, decreasing the variance with a factor $10$ further without increasing the sample size. We give numerical results in Table~\ref{tab:LPMimportance1} and present a schematic view of how IS and LPM are combined in Figure~\ref{fig:sampleschematic}.

\begin{figure}[h!]
\begin{subfigure}{.49\textwidth}
  \centering
  \includegraphics[width=1.0\linewidth]{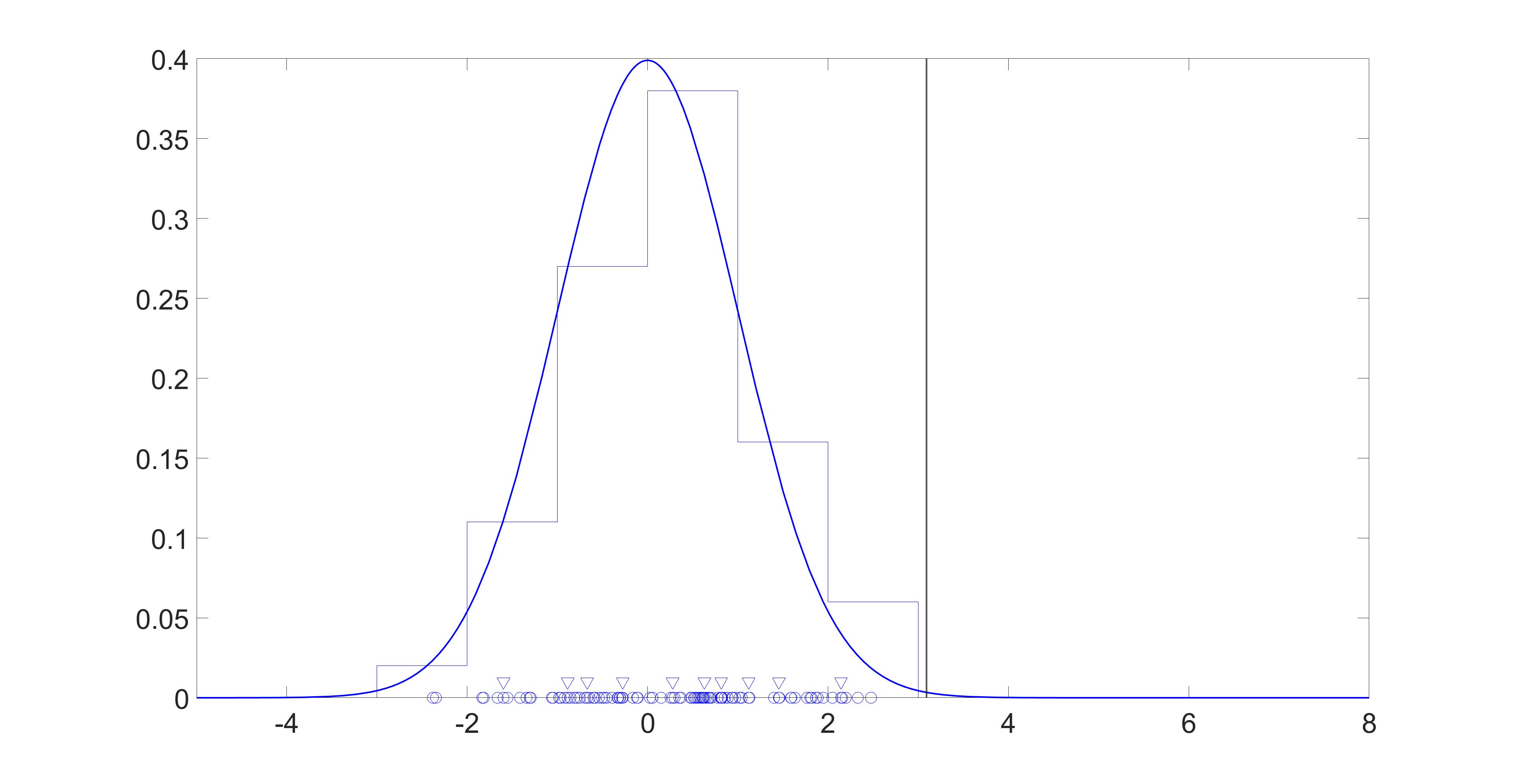}
  \caption{Discretizing distribution $\mathcal N(0,1)$.}
\end{subfigure} \hfill
\begin{subfigure}{.49\textwidth}
  \centering
  \includegraphics[width=1.0\linewidth]{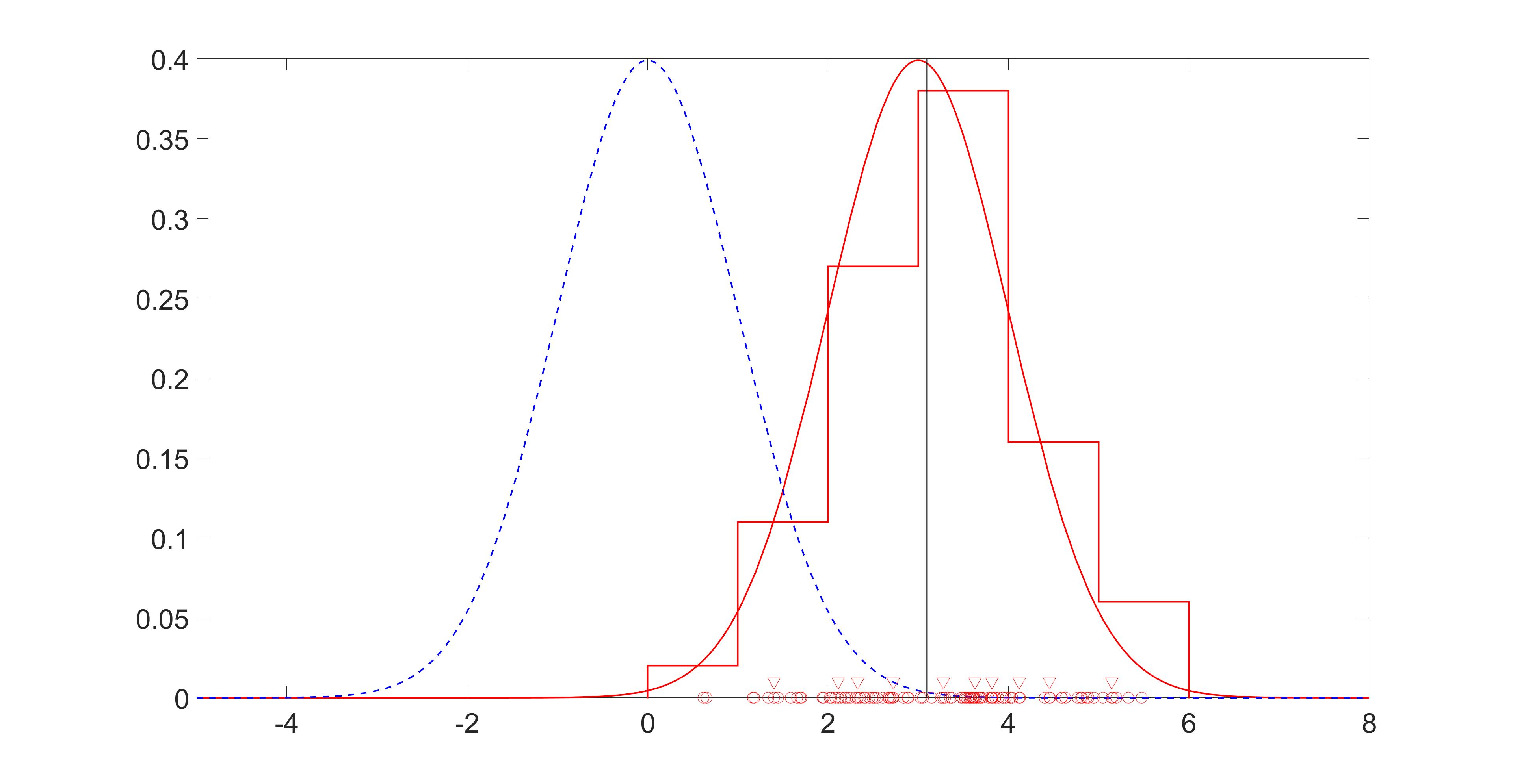}
  \caption{Discretizing distribution $\mathcal N(3,1)$.}
\end{subfigure}
\caption{Substantially more of the sample points end up in the non-zero region after shifting the discretizing distribution. Circles ($\circ$) represents the discretizing iid sample ($N=100$), triangles ($\nabla$) the LPM thinning ($n=10$), and the staircase line is a normalized histogram of the discretizing sample. The black vertical line denotes $\alpha_{0.999}$.}
\label{fig:sampleschematic}
\end{figure}

\begin{table}[h!]
    \centering
    \begin{tabular}{|c|c|c|c|c|c|c|c|c|}
\hline
& $\hat \mu_{iid}$  & $\hat \sigma_{iid}$&  $\hat \mu_{LPM}$ & $\hat \sigma_{LPM}$ & $\hat \mu_{IS}$ & $\hat \sigma_{IS}$   & $\hat \mu_{IS \& LPM}$ & $\hat \sigma_{IS \& LPM}$  \\ \hline
    $ n= 10^2$  &   0     & 11.101  &    0     & 10.601   &  2.663 & 0.614  &  3.281&0.147 \\
    $ n= 10^3$  &   6.424 & 3.352    &   3.094& 2.043     &  3.332 & 0.191   &  3.412 &0.063  \\
    $n=  10^4$  &   1.953 & 1.097    &   -        &   -            &  3.391 &0.062    &    -    &   -  \\  \hline
    \end{tabular}
    \caption{Numerical estimate of  $\mu = \mathbb E[y_{0.999}(X)] \approx  3.367$ using iid, LPM2, IS, and a combination of IS and LPM2. Combining LPM with IS gives a reduction of a factor 10 in the variance of the estimate. The discretization for LPM is done with $N=10^4$ points.} 
    \label{tab:LPMimportance1}
\end{table}

IS in combination with stratification has shown to be extremely efficient in the context of financial mathematics, see \cite{G04}, and this result is clearly replicated for LPM. We stress that LPM requires no adaptations, is applied with a single line of code, and gives approximately the same variance with only $\frac{1}{10}$ of the sample points compared to using only IS.
\end{example}


\section{Example applications}

 \label{sec:applications}

There are obviously numerous possible applications of the LPM on continuous distributions; in this section we give a few examples from mathematical finance (option pricing) and dynamical systems (estimating non-local stability). 

\subsection{Mathematical finance: Pricing a European option} \label{sec:EC}

A European call option (EC) is a financial derivative which gives its holder the right but not the obligation to buy a certain asset $S$, at time $t=T$, for the price $K$. The asset $S$ is usually called the underlying of the contract, $T$ the maturity, and $K$ the strike price. It is a standard result of financial mathematics that the fair price of such a contract is given by its discounted expected value under the so called \textit{risk neutral measure}. In fact, the characterization of the ``fair price'' as an expected value under a certain measure is not specific to European options but holds in general. Therefore, simulation and variance reduction are common tools for estimating prices of financial derivatives.

In the special case of a EC written on an underlying $S$ which evolves as a geometric Browninan motion,
\begin{equation} \label{eq:GBM}
dS_t = \mu S_t dt + \sigma S_t dW_t, \qquad S_0=s,
\end{equation}
this price is given by $
P_{EC} =  \mathbb E \left [ \max(0, S_T-K) \right]$ where
$S_T= s \exp( (r-\frac{1}{2}\sigma^2)T + \sigma \sqrt{T} Z)$ , $Z \sim \mathcal N(0,1)$.
In this particular setting an exact solution is available,
\begin{equation} \label{eq:pricetheoretical}
P_{EC}(s)= N(d_1) s - N(d_2) K e^{-rT}, \qquad \mbox{where}
\end{equation}
\begin{equation*}
d_1= \frac{1}{\sigma \sqrt{T-t}} \left [\log\left( \frac{s}{K}\right) + \left ( r+ \frac{\sigma^2}{2}\right)(T-t) \right]\qquad \mbox{and} \qquad d_2 = d_1 -\sigma \sqrt{T-t},  \\
\end{equation*}
making the EC well suited for testing simulation techniques and variance reduction methods.

Here we consider a EC with $s = 100$, $r = 0.03$, $\sigma = \frac{1}{2}$, $T = 1/4$, and  $K = 120$ for which the theoretical price given by \eqref{eq:pricetheoretical} is $P_{EC}\approx 3.886$. Estimated prices based on simulation with and without variance reduction through LPM are given in Table \ref{tab:LPMEC}. A substantial variance reduction is achieved with LPM.

The variance $\hat \sigma_{LPM}$ is calculated naively by repeating the experiment $m=10^4$ times and $\hat V( \hat P_{EC}^{LPM})$ is calculated as outlined in Section \ref{sec:varianceest} using the $10$ nearest neighbours. Figure~\ref{fig:ECdecrease} shows how the variance of the estimate declines as a function of the discretization parameter $N$ for some fixed values of the sample size $n$.
\begin{table}[h!]
    \centering
    \begin{tabular}{|c|c|c|c|c|c|}
\hline         & $\hat P_{EC}^{iid}$  & $\hat \sigma_{iid}$& $\hat P_{EC}^{LPM}$ & $\hat\sigma_{LPM}$ & $\sqrt {\hat V_{LPM}(\hat P_{EC}^{LPM})}$\\ \hline
    $n=10^2$  &  3.523 &   0.899    &  4.089  &  0.307  & 0.374   \\
    $n= 10^3$  & 3.748 &   0.365    &  3.899  &  0.116  & 0.043  \\ \hline    %
    \end{tabular}
    \caption{Estimates of $P_{EC}\approx 3.886$ using iid normal random numbers and LPM2 samples. The discretization for LPM is done with $N=10^4$ points. The last column refers to the variance estimate of Section \ref{sec:varianceest}}
    \label{tab:LPMEC}
\end{table}

%

\subsection{Nonlinear dynamics: Estimating non-local stability}

%


A dynamical system (DS) is a set of differential equations governing the time dependence of a point. Examples include the motion of a falling object, the swinging of a clock pendulum, the flow of water in a pipe, and the number of fish in a lake.
A DS can be written on the form
\begin{align}\label{eq:dyndef}
\frac{dx}{dt} &= f (x, t), \quad \quad x = x(t) \in \mathbf{R}^d, \quad t > 0,
\end{align}
where $f (x, t)$ is a given function defined on $\mathbf{R}^d \times \mathbf{R}$ taking values in $\mathbf{R}^d$ and $d$ is the space dimension. 
Initially, at $t = 0$, $x$ is given by an initial condition
$$
x(0) = (x_1(0), x_2(0), \dots, x_d(0)) \in \mathbf{R}^d.
$$
Understanding the stability of solutions of a DS is very important for many applications.
A simple way of quantifying the stability is to test the solutions ability to withstand perturbations.
For nonlinear DSs such investigation naturally split into local and non-local analysis.
The local stability approach considers small perturbations, is usually based on linearizations and yields
information only in a small neighborhood of the solution.
The non-local approach we consider here considers also large perturbations and thereby involves investigations of the \emph{basin of attraction} for the solution of the DS, see e.g. \cite{TNM06,MHMK13,L18}.
Indeed, we can estimate the (non-local) stability of an operating electric hydro power generator (as in the second example below) by first modelling it by a DS,  then solving the DS and thereby finding a solution representing the operating state, and finally repeatedly test if the DS recovers the same solution after a given perturbation. If the solution is recovered for a large set of perturbations then we say it is stable.
%
%
A perturbation can naturally be modelled through the systems initial condition. Therefore, to understand the stability of a DS one solves the equations from a set of different initial conditions and study the resulting behaviour.
%
We call an initial condition \textit{safe} if it takes the system to the desired solution after a reasonable time limit and unsafe otherwise. Let $\mathbb{N}_{\text{safe}}$ and $\mathbb{N}_{\text{tot}}$ denote the number of the
safe and tested initial conditions, respectively.
We obtain a simple stability measure by
\begin{align}\label{eq:def_P}
\mathcal{P} = \frac{\mathbb{N}_{\text{safe}}}{\mathbb{N}_{\text{tot}}}
\end{align}
as the fraction of safe initial conditions.
Measure $\mathcal{P}$ has been considered in e.g. \cite{LA07,MHMK13,L18} to which we also refer the reader for further discussions and applications on non-local stability measures.
What remains then is to choose a suitable set of initial conditions (perturbations) to test the DS in \eqref{eq:dyndef} for.
These perturbations can be taken deterministically or randomly from a predefined probability distribution; the choice should reflects what the system may be exposed to in reality and is therefore case specific. 

Since the system of equations \eqref{eq:dyndef} may require substantial computational power and be time consuming, it has a value to implement variance reduction whenever initial conditions are randomly sampled.
Let us also mention that, in addition to stability, the related concept of resilience obeys similar application for variance reduction,
see e.g. \cite{MKD15,LLMBB19, L18} for constructions and applications of non-local resilience measures.
We proceed with two examples, the first origins from \cite{MHMK13} and considers a rainforest,
while the second expands on \cite{LA07} and investigates non-local stability in an electric generator.

\subsubsection{A simple model of a rainforest}
The Amazon rainforest may be assumed to have two stable states: a fertile forest state and a barren savanna state. This dual stability of the Amazonas (also called \textit{bistability}) arises from a positive feedback loop occuring in the rain forest: Deep-rooting trees take up water stored in the soil and transpire it to the atmosphere. Overall precipitation therefore increases in forest covered areas and a rather arid area may still be supportive of forest growth if its forest cover exceeds a certain critical threshold. On the other hand, if the forest cover goes below this threshold the area would lose all of its trees.
This behaviour can be summarized in the following simple DS,
\begin{align*}
\frac{d x}{d t} = F(x) -  M x, \quad \text{where} \quad F(x) = \left\{
\begin{array}{ll}
R x (1-x) \quad &\text{if} \quad x > x_{crit},\\
0 \quad  &\text{if} \quad x \leq x_{crit}.
\end{array}
\right.
\end{align*}
Here, $x = x(t)$ is the relative forest cover, $R$ gives the growth rate, $M$ the death rate, and $x_{crit}$ being the critical forest cover threshold.
This model's two equilibria are the forest state $x_F$ and the savanna state $x_S$, given by
$$
x_{F} = 1 - \frac{M}{R} \quad \mbox{and} \quad	x_S=0,
$$
respectively. Both equilibria exist and are stable if $x_F > x_{crit}$.


We test our model with perturbations corresponding to a normally distributed decrease in forest biomass and therefore we sample $n$ points $\{z^1, z^2, \dots, z^n\}$ from an $\mathcal N(0,1)$ distribution and produce $n$ different initial conditions as
$$
x^i(0)= x_F-|z^i|, \quad i = 1,2,\dots, n.
$$
To numerically integrate trajectories from their initial conditions, we used MATLAB’s ode-solver ODE45 with standard tolerance settings. We integrate each trajectory until the solution trajectory $x(t)$ reaches one of the two neighborhoods
$$
(x_F - \epsilon, x_F + \epsilon) \quad \mbox{or} \quad (x_S - \epsilon, x_S + \epsilon),
$$
representing the forest equilibrium and the savanna equilibrium, respectively.

In Figure~\ref{fig:rainforest} we present simulations of the non-local stability measure
$\mathcal{P}$, counting the fraction of safe initial conditions as defined in \eqref{eq:def_P},
with and without using variance reduction through LPM.
We have sampled $n = 50$ initial conditions with LPM (for each value of $x_{crit}$) using $N = 10^4$ points in the initial iid discretization of the normal distribution.
A substantial variance reduction can clearly be observed.
We also see that the stability measure $\mathcal{P}$ decreases rapidly as the critical threshold $x_{crit}$ approaches the bifurcation value at 0.5.
This means that even though the forest can survive (that is, $x_{F}$ exists), it is very unlikely due to the fact that in reality, small perturbations are likely present.

\subsubsection{An electric hydropower generator}
Large synchronous electric generators usually have small air-gaps between the rotor and the stator.
Normally, this gap is about 0.2$\%$ of the stator radius. Measurements on generators indicate asymmetry in this air-gap due to, e.g., production imperfections. These asymmetries distort the magnetic flux density in the gap, resulting in an attraction force between the rotor and the stator, usually called unbalanced magnetic pull (UMP). The effect of UMP can be vibrations which may eventually be dangerous to the machine. The following DS was derived in \cite{LA07} as a dimensionless version of the equations of motion for a hydropower generator: 
\begin{align}\label{eq:huvudsyst}
X''\, +\, 2\, \zeta\, X'\, +\, X\,
=  \,F_X\left(X,Y,\tau\right),\nonumber\\
Y''\, +\, 2\, \zeta\, Y'\, +\, Y\,
=  \,F_Y\left(X,Y,\tau\right).
\end{align}
%
%
%
Here, $X$ and $Y$ give the location of the rotor center, $\zeta = 0.1$ is a damping ratio and $\tau$ is a dimensionless time.
Physical considerations yield the forces $F_X$ and $F_Y$ as
\begin{align}\label{eq:fxxx}
F_X\, &=\,  \frac{k_m}{2\, \pi\, k}\, \int_0^{2\pi}\, \frac{\cos \varphi}{G\left(X,Y,\tau,\varphi\right)^2}\, d\varphi,\notag\\
F_Y\, &=\, \frac{k_m}{2\, \pi\, k}\, \int_0^{2\pi}\, \frac{\sin \varphi}{G\left(X,Y,\tau,\varphi\right)^2}\, d\varphi,
\end{align}
in which $k \approx 3.456\cdot10^8 N/m$ is mechanical stiffness, $k_m \approx 1.4715\cdot10^8 N/m$ is electromagnetic stiffness,
and the air-gap is given by
\begin{align}\label{eq:gapet}
G\,& =\, 1\, + \, \left(\Delta\, -\, X\right)\,\cos\varphi\, -\, Y\,\sin\varphi.
\end{align}
The parameter $\Delta$ models imperfections in the generator by placing the rotor center a distance $\Delta$ from the stator center when the machine is at rest (giving asymmetry in the air-gap resulting in UMP). This kind of asymmetry has been extensively studied for generators and is known as \textit{eccentricity}.

Considering the velocities $X'$ and $Y'$ as additional dependent variables allows us to convert the second order
two dimensional DS in \eqref{eq:huvudsyst} into a four dimensional DS on the form \eqref{eq:dyndef},
describing the motion of the rotor through $(X(\tau), Y(\tau), X'(\tau), Y'(\tau))$, for $\tau > 0$.
System \eqref{eq:huvudsyst} obeys a stable equilibrium $E_\Delta$ as long as the eccentricity $\Delta$ is small enough.
When $\Delta = 0$ the machine is perfect and we have $E_0 = (0,0,0,0)$. As $\Delta$ increases, $E_\Delta$ moves away from the origin and the air-gap becomes asymmetric.
At a certain value of $\Delta$, the equilibrium $E_\Delta$ disappears in a so called fold bifurcation, see \cite{LA07}.
In addition to the equilibrium $E_\Delta$, system \eqref{eq:huvudsyst} always has the rotor-stator contact state when the air-gap becomes non-positive,
corresponding to a complete failure of the machine.

In case of the generator under study, it is natural to consider displacement of the rotor center, velocity
impulses on the rotor, or combinations of them as perturbations.
A velocity impulse may be due to shock of the rotor, while a displacement may be due to a shock of the bed-plate of the machine.
We here chose to test the machine for perturbations normally distributed in both displacement and velocity,
giving us a $4$-dimensional space of perturbations (displacement in $X$ and $Y$ and velocity in $X$- and $Y$-direction).
We chose to sample $n$ points from the $4$-dimensional normal distribution centered at equilibrium $E_\Delta$ with
standard deviation $\sigma=0.4$ in all four directions.
In particular, we sample $n$ points
$$
\{z^1, z^2, \cdots, z^n\} \quad \mbox{where} \quad z^i= (z^i_{1}, z^i_{2}, z^i_{3}, z^i_{4} ) \sim \mathcal N^4(E_\Delta,\sigma)
$$
and define our $n$ initial conditions as
$$
(X^i(0), Y^i(0), X'^{\,i}(0), Y'^{\,i}(0)) = E_\Delta +  (z^i_{1}, |z^i_{2}|, z^i_{3}, z^i_{4} ),\quad \text{for} \quad i = 1,2,\dots, n.
$$
Here, the modulus on $z^i_{2}$ cuts the sample space in two halves. This is possible due to symmetry in the generator model;
we may consider only the halfspace $Y\geq 0$ and therefore we mirror our initial conditions to this set.
To numerically integrate trajectories from
their initial conditions to the attractor, we used MATLAB’s ode-solver ODE45
with standard tolerance settings.
We integrate each trajectory until it reaches the small neighborhood given by a ball of radius $0.01$ centered at $E_\Delta$,
or to the first point when the air-gap is nonpositive, representing stable operation and complete machine failure, respectively.

\begin{figure}
\begin{subfigure}{.49\textwidth}
  \centering
  \includegraphics[width=7.5cm, height=6.5cm]{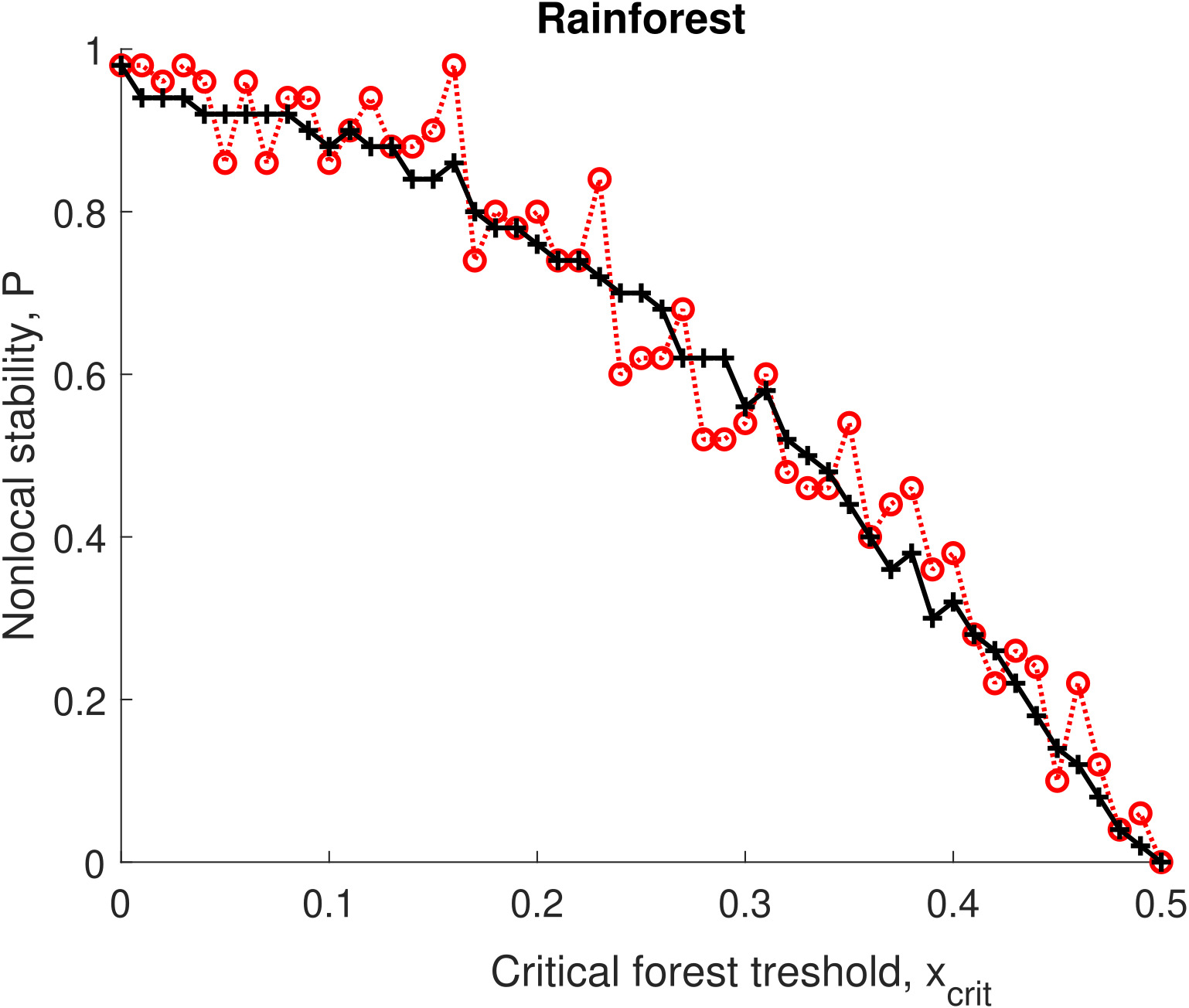}
  \caption{The measure $\mathcal{P}$ as function of the
critical forest threshold $x_{crit}$, with and without variance reduction.
50 initial conditions are examined for each value of $x_{crit}$. Parameters are set to $R = 1$ and $M = 1/2$.} 
\label{fig:rainforest}\end{subfigure} \hfill
\begin{subfigure}{.49\textwidth}
  \centering
  \includegraphics[width=7.5cm, height = 6.6cm]{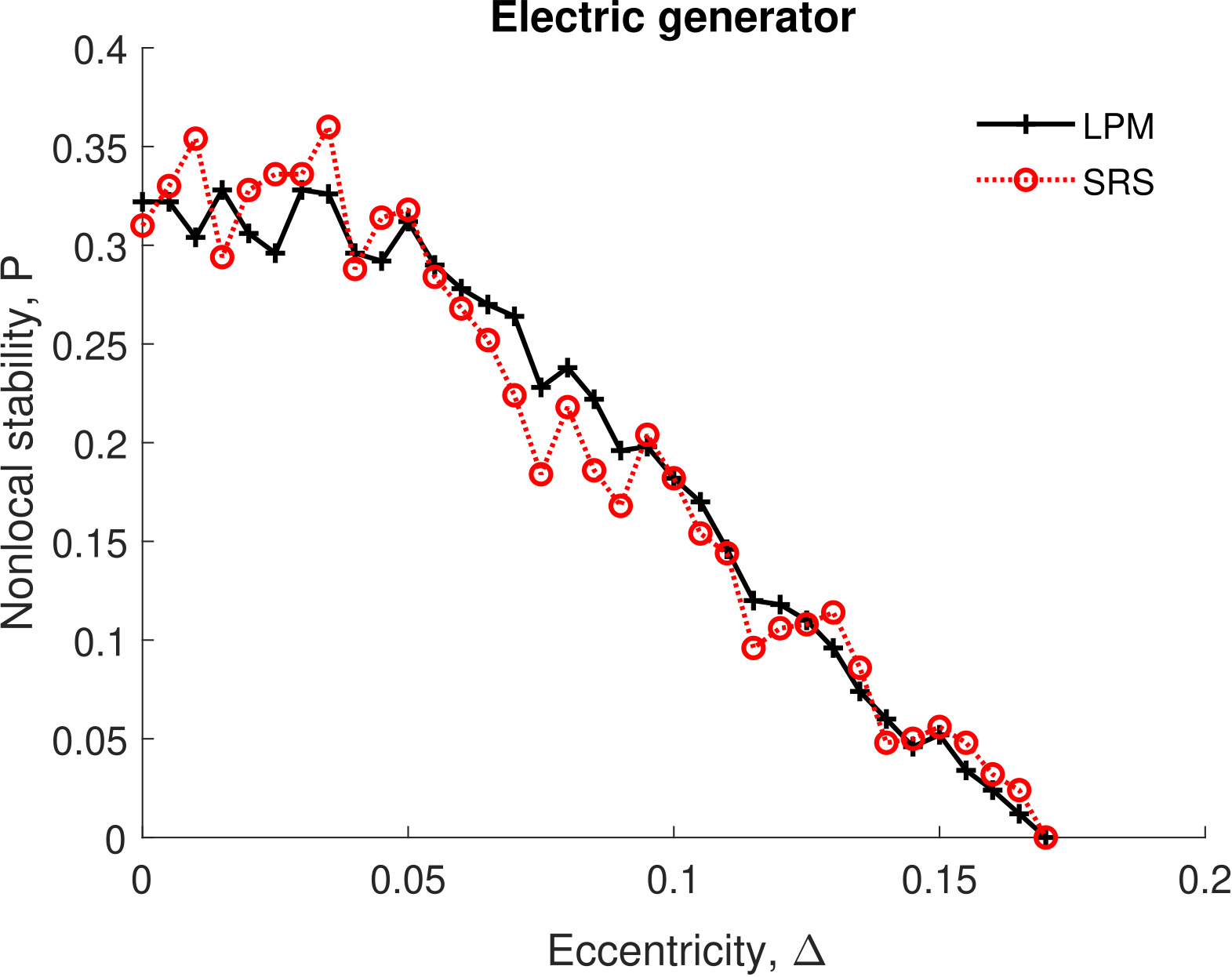}
  \caption{The measure $\mathcal{P}$ as functions of the eccentricity $\Delta$ with and without variance reduction. 500 initial conditions are examined for each value of $\Delta$.
Each data point needed several days to complete.} 
\label{fig:rotor}\end{subfigure}
\caption{Stability of the forest equilibrium in the rainforest model (left) and the operating equilibrium in the generator model (right).}
\label{fig:decrease}
\end{figure}

In Figure~\ref{fig:rotor} we present simulations of the non-local stability measure $\mathcal{P}$, counting the fraction of safe initial conditions as defined in \eqref{eq:def_P}, with and without using variance reduction through LPM.
We perform our simulation with an LPM sample of $n = 500$ initial conditions (for each value of the eccentricity $\Delta$) using $N = 10^4$ points in the initial iid discretization. As in the rainforest model discussed above, a reduction in the variance can be observed.

We remark that the forces $F_X$ and $F_Y$ in \eqref{eq:fxxx} are given as integrals which have to be evaluated at each time step when numerically solving the DS \eqref{eq:huvudsyst}. This makes the solution procedure very time consuming and thus, when randomly inferring perturbations, a robust variance reduction method as suggested here is clearly motivated as it can significantly decrease the cost of the stability investigation.


\section{Variance estimation} \label{sec:varianceest}
With many variance reduction techniques, a variance reduced estimate is often at odds with the possibility to estimate this variance. The variance of the most naive estimate $\hat \mu_{iid}$ of $\mu=\mathbb E(y(X))$, based on $n$ independent observations of the random variable $X$, 
can easily be estimated via the sample variance as
$$
\hat V (\hat \mu_{iid}) =
\frac{1}{n(n-1)} \sum_{x \in S} \left(y(x) - \frac{1}{n}\sum_{x' \in S}y(x')\right)^2.
$$
When variance reduction techniques are applied, the estimation of the variance is usually more complicated.
%
%
For LPM, we suggest to estimate the variance using a local mean variance estimator as in \cite{SO03, SO04} and \cite{GS14}, see also \cite{GM18}. More precisely, we suggest to estimate the variance of our LPM-estimator of $\mathbb E(y(X))$ by the local mean variance estimator 
\begin{equation}\label{eq:varianceest}
\hat V (\hat \mu_{LPM}) = 
\frac{n'}{n^2 (n'-1)} \sum_{x \in S} \left(y(x) - \frac{1}{n'}\sum_{x' \in S_{x}}y(x')\right)^2.
\end{equation}
where $n'$ is the size of a local neighbourhood $S_{x} \subset S$ of point $x$, consisting of point $x$ and its $n'-1$ nearest neighbours. Note that, with $n'=n$, the estimator $\hat{\sigma}^2_{LM}$ is identical to $\hat{\sigma}^2_{iid}$. The estimator \eqref{eq:varianceest} clearly depends on $n'$ and this parameter needs to be chosen with care. The optimal $n'$ and the performance of the suggested variance estimator calls for a thorough investigation outside the scope of this paper; we simply refer the reader to the estimates in Table \ref{tab:LPMEC} which are calculated with $n'=10$.

\section{Quick start tutorial}

For the readers convenience, we here provide a quick start guide to using LPM in R and MATLAB. 

\subsection{R}

The LPM is implemented in two different version in R, \texttt{lpm1} and \texttt{lpm2}, both using Euclidean distance. The implementations are found in the package \texttt{BalancedSampling}. 
The syntax is as follows:
$$
\texttt{s=lpm1(prob,X)} \quad \mbox{or}  \quad \texttt{s=lpm2(prob,X)}
$$
where
\begin{itemize}
\itemsep0em
\item[] \texttt{prob} - vector of length N with inclusion probabilities
\item[]\texttt{X} - $N \times q$ matrix representing the discrete population
\item[] \texttt{s} - row numbers of $X$ constituting a balanced LPM sample.
\end{itemize}
An explicit example is the generation of Figure \ref{fig:2Dunif} a), which is done by the code below. We stress that the only difference from generating Figure \ref{fig:2Dunif} b) is line 6, "select LPM-sample".
\begin{align*}
&\texttt{library(BalancedSampling); library(deldir)} &&\mbox{\# import libraries} \\
&\texttt{set.seed(1);} &&\mbox{\# set seed and sample-size} \\
&\texttt{N = 10000; n = 100;} &&\mbox{\# set sample size} \\
&\texttt{p = rep(n/N,N);} &&\mbox{\# set (equal) incl. prob.}\\
&\texttt{X = cbind(runif(N),runif(N));} &&\mbox{\# discretize population}\\
&\texttt{s = lpm2(p,X); X=X[s,];} &&\mbox{\# \textbf{select LPM-sample}}\\
&\texttt{LPMtesselation = deldir(X[1:n,])} &&\mbox{\# create tesselation (deldir-pack.)}\\
&\texttt{LPMtiles = tile.list(LPMtesselation)} &&\mbox{}\\
&\texttt{plot(LPMtiles, pch=19)} &&\mbox{\# plot figure}
\end{align*}



\subsection{MATLAB}

In MATLAB, an implementation of LPM2 with arbitrary distance function together with a variance estimate is available. To reduce computational speed, this implementation is rather memory demanding and is best used for discrete populations of size $N=10^4$ or smaller. The syntax is as follows:
$$
\texttt{[s,svar]=lpm2(prob,X,distfcn,ns)}
$$
where
\begin{itemize}
\itemsep0em
\item[] \texttt{prob} - vector of length N with inclusion probabilities
\item[] \texttt{X} - $N \times q$ matrix representing the discrete population
\item[] \texttt{distfcn} - distance function. All matlab standard distances are available, e.g., 'euclidean', 'cityblock', or 'chebychev'. User specified distance is also possible, see Remark \ref{remark:distancefunction}.
\item[] \texttt{ns} - a positive integer giving the number of nearest neighbours (including the point itself) to be used in the variance estimate. The default value is \texttt{ns=1} and variance estimation is then ignored.
\item[] \texttt{s} - row numbers of $X$ constituting a balanced LPM sample.
\item[] \texttt{svar} - $M \times (ns-1)$ matrix where $M=length(s)$ and 
the elements in row $i$ are the row numbers of $X$ representing the nearest neighbours to \texttt{s(i)} (which are also in \texttt{s}).
\end{itemize}

\begin{remark} \label{remark:distancefunction}
Any user specified function is allowed in the MATLAB function \texttt{lpm2}; \texttt{distfcn} should then be the string \texttt{@fcnhandle}, where \texttt{fcnhandle} is a user-specified function taking as arguments a $1 \times d$ vector $X_i$ containing a single row of $X$, an $M\times d$ matrix $\tilde X$ containing multiple rows of $X$, and returning an $M \times 1$ vector of distances, whose $j$th element is the distance between the observations $X_i$ and $\tilde X(j,:)$.
\end{remark}

An explicit example in MATLAB is the calculation of $\hat P_{EC}^{LPM}$ in Table \ref{tab:LPMEC}, which is done by the code below. We stress that, again, the only difference between the variance reduced LPM-estimate and that based on iid observations is a single line of code, line 5, "select LPM-sample".
\begin{align*}
&\texttt{rng(1); N = 10000; n = 100;} &&\hspace{-3cm }\mbox{\% set seed and sample size} \\
&\texttt{p=ones(N,1)*n/N;} &&\hspace{-3cm }\mbox{\% set (equal) incl. prob.}\\
&\texttt{S0=100; K=120; T=1/4; r=.03; sigma=1/2;} &&\hspace{-3cm }\mbox{\% problem parameters}\\
&\texttt{X=randn(N,1);} &&\hspace{-3cm }\mbox{\% {discretize population}}\\
&\texttt{s=lpm2(p,X,'euclidean'); X=X(s); p=p(s);} &&\hspace{-3cm }\mbox{\% \textbf{select LPM-sample}}\\
&\texttt{stockvalue= S0*exp((r-1/2*sigma\^{}2)*T + sigma*sqrt(T).*X(1:n));} &&\mbox{}\\
&\texttt{optionpayoff = exp(-r*T)*max( 0, stockvalue - K );} &&\hspace{-3cm }\mbox{\% stock and option value} \\
&\texttt{priceest=sum(optionpayoff./p(1:n))/N} &&\hspace{-3cm }\mbox{\%final price estimate}
\end{align*}


\begin{thebibliography}{99}


\bibitem{AKB23}
Allard, A., Keskitalo, E.C.H., \& Brown, A. (Eds.). (2023). {\it Monitoring Biodiversity: Combining Environmental and Social Data} (1st ed.). Routledge. 

%
\bibitem{BPP15} Benedetti, R., Piersimoni, F., \& Postiglione, P. (2015). {\it Sampling spatial units for agricultural surveys}. Berlin: Springer.

%
\bibitem{B21}
Brus, D. J. (2021). Statistical approaches for spatial sample survey: Persistent misconceptions and new developments. {\it European Journal of Soil Science}, 72(2), 686-703.


\bibitem{DT98}
Deville, J-C, Tille, Y. (1998)
{\it Unequal probability sampling without replacement through a splitting method}, Biometrika 85(1), 89-101.

\bibitem{G04}
Glasserman P.,
{\it Monte Carlo methods in financial engineering},
 Vol. 53. New York: springer, 2004.

\bibitem{Gea17a}
Grafström, A., Zhao, X., Nylander, M., \& Petersson, H. (2017). A new sampling strategy for forest inventories applied to the temporary clusters of the Swedish national forest inventory. {\it Canadian Journal of Forest Research}, 47(9), 1161-1167.

\bibitem{GM18}
Grafström, A., \& Matei, A. (2018). Spatially balanced sampling of continuous populations. {\it Scandinavian Journal of Statistics}, 45(3), 792-805.








\bibitem{Gea17b}
Grafström, A., Schnell, S., Saarela, S., Hubbell, S. P., \& Condit, R. (2017). The continuous population approach to forest inventories and use of information in the design. {\it Environmetrics}, 28(8), e2480.

\bibitem{GLP22}
Grafström, A., Lisic, J., Prentius, W. (2022). BalancedSampling: Balanced and Spatially Balanced
Sampling. R package version 1.6.3, URL https:\slash\slash CRAN.R-project.org\slash package=BalancedSampling.



\bibitem{GLS12}
Grafstr\"om A., Lundstr\"om N.L.P., Schelin L.,
{Spatially balanced sampling through the pivotal method},
{\it Biometrics} 68.2 (2012): 514--520.

\bibitem{GL13}
Grafstr\"om A., Lundstr\"om N.L.P.,
{Why well spread probability samples are balanced},
{\it Open Journal of Statistics} 3.1 (2013): 36--41.


\bibitem{GS14}
Grafström, A., \& Schelin, L. (2014). How to select representative samples. {\it Scandinavian Journal of Statistics}, 41(2), 277-290.

\bibitem{HT52}
Horvitz, D. G., \& Thompson, D. J. (1952). A generalization of sampling without replacement from a finite universe. {\it Journal of the American statistical Association}, 47(260), 663-685.


\bibitem{L18}
Lundstr\"om N.L.P.,
{How to find simple non-local stability and resilience measures},
{\it Nonlinear dynamics} 93.2 (2018): 887--908.

\bibitem{LA07}
Lundstr\"om N.L.P., Aidanp\"a\"a.,
{Dynamic consequences of electromagnetic pull due to deviations in generator shape},
{\it Journal of sound and vibration} 301.1--2 (2007): 207--225.

\bibitem{LLMBB19}
Lundstr\"om N.L.P., Loeuille N., Meng X., Bodin M.,  Br\"annstr\"om, {\AA}.
{Meeting yield and conservation objectives by harvesting both juveniles and adults},
{\it The American Naturalist}, 193(3), (2019): 373--390.

\bibitem{MHMK13}
Menck P.J., Heitzig J., Marwan N., Kurths, J.),
{How basin stability complements the linear-stability paradigm},
{\it Nature physics}, 9(2), (2013): 89--92.

\bibitem{MKD15}
Mitra C., Kurths J.,  Donner R. V.
{An integrative quantifier of multistability in complex systems based on ecological resilience},
{\it Scientific reports}, 5(1), (2015): 1--10.


\bibitem{SO03}
Stevens Jr, D. L., \& Olsen, A. R. (2003). Variance estimation for spatially balanced samples of environmental resources. {\it Environmetrics}, 14(6), 593-610.

\bibitem{SO04}
Stevens Jr, D. L., \& Olsen, A. R. (2004). Spatially balanced sampling of natural resources. {\it Journal of the American statistical Association}, 99(465), 262-278.


\bibitem{TNM06}
Tan Y., Nešić D., Mareels I.,
{On non-local stability properties of extremum seeking control}.
{\it Automatica} 42.6 (2006): 889--903.




%
%




\end{thebibliography}
\end{document}